\documentclass[preprint,nofootinbib,aps,superscriptaddress]{revtex4}

\usepackage{graphicx}

\newcommand{\beq}{\begin{equation}}
\newcommand{\eeq}{\end{equation}}
\newcommand{\beqa}{\begin{eqnarray}}
\newcommand{\eeqa}{\end{eqnarray}}
\newcommand{\nn}{\nonumber}

\def\OMIT#1{{}}
\def\lqcd{\Lambda_{\rm QCD}}
\def\d{{\rm d}}

\def\bra#1{\langle#1|}
\def\ket#1{|#1\rangle}

\def\Fb{{\cal F}_\Lambda}

\def\wmax{w_{\rm max}}

\newcommand{\Bbar}{\,\overline{\!B}}
\def\B0bar{\Bbar{}^0}

\newcommand{\SCETa}{SCET$_{\rm I}$}
\newcommand{\SCETb}{SCET$_{\rm II}$}

\def\nslash{n\!\!\!/}

\usepackage[usenames]{color}    

\begin{document}

\preprint{\vbox{\hbox{LBNL--54153} \hbox{MIT-CTP-3453} 
  \hbox{CALT-68-2467} \hbox{hep-ph/0312319}}}

\vspace*{1.25cm}

\title{\boldmath Predictions for nonleptonic $\Lambda_b$ and $\Theta_b$
decays\vspace{4pt}}

\author{Adam K.\ Leibovich}\email{akl2@pitt.edu}
\affiliation{Department of Physics and Astronomy, University of Pittsburgh,
        \\[-5pt] Pittsburgh PA 15260}
        
\author{Zoltan Ligeti}\email{ligeti@lbl.gov}
\affiliation{Ernest Orlando Lawrence Berkeley National Laboratory,\\[-5pt]
        University of California, Berkeley CA 94720}

\author{Iain W.\ Stewart\,}\email{iains@mit.edu}
\affiliation{Center for Theoretical Physics, Massachusetts Institute for
        Technology,\\[-5pt] Cambridge, MA 02139}

\author{Mark B.\ Wise\,\vspace{10pt}}\email{wise@theory.caltech.edu}
\affiliation{California Institute of Technology, Pasadena, CA 91125
        \\ $\phantom{}$ }


\begin{abstract}

We study nonleptonic $\Lambda_b \to \Lambda_c\pi$, $\Sigma_c\pi$ and
$\Sigma_c^*\pi$ decays in the limit $m_b, m_c, E_{\pi}\gg \lqcd$ using the
soft-collinear effective theory.  Here $\Sigma_c=\Sigma_c(2455)$ and
$\Sigma_c^* = \Sigma_c(2520)$. At leading order the $\Lambda_b \to
\Sigma_c^{(*)}\pi$ rates vanish, while the $\Lambda_b \to \Lambda_c\pi$ rate is
related to $\Lambda_b \to \Lambda_c\ell\bar\nu$, and is expected to be larger than
$\Gamma(B\to D^{(*)}\pi)$.  The dominant contributions to the $\Lambda_b \to
\Sigma_c^{(*)}\pi$ rates are suppressed by $\lqcd^2/E_{\pi}^2$.  We predict
$\Gamma(\Lambda_b \to \Sigma_c^*\pi) \,/\, \Gamma(\Lambda_b \to \Sigma_c\pi) =
2 + {\cal O}[\lqcd/m_Q,\, \alpha_s(m_Q)]$, and the same ratio for
$\Lambda_b\to \Sigma_c^{(*)} \rho$ and for $\Lambda_b\to \Xi_c^{(\prime,*)}K$. 
``Bow tie'' diagrams are shown to be suppressed. We comment on possible
discovery channels for weakly decaying pentaquarks, $\Theta_{b,c}$ and their
nearby heavy quark spin symmetry partners, $\Theta_{b,c}^*$.

\end{abstract}

\maketitle

Heavy baryon decays are interesting for many reasons.  Heavy quark
symmetry~\cite{Isgur:ed} is more predictive in semileptonic $\Lambda_b \to
\Lambda_c \ell\bar\nu$ decay  than in $B\to D^{(*)} \ell\bar\nu$, and may
eventually give a determination of $|V_{cb}|$ competitive with meson
decays~\cite{Isgur:pm}.  In this paper we concentrate on the more complicated
case of nonleptonic $b\to c\bar u d$ baryon transitions, as shown in
Table~\ref{tab:charm}. These channels provide a testing ground for our
understanding of QCD in nonleptonic decays.  Our analysis is based on heavy
quark symmetry and the soft-collinear effective theory
(SCET)~\cite{Bauer:2000ew}.

\begin{table}[b!]
\tabcolsep 8pt
\begin{tabular}{cccclc}
\hline\hline
Notation  &  $s_l$  &  $I(J^P)$  &  Mass (MeV) &  Decays considered
\\ \hline 
$\Lambda_c$, $\Lambda_b$      &  $0$  &  $0(\frac12^+)$  &  2285, 5624 &
 $\Lambda_b \to \Lambda_c^+ \pi^-$
 \\
$\Sigma_c=\Sigma_c(2455)$   &  $1$   &  $1(\frac12^+)$  &  2452 &
 $\Lambda_b \to$ $\Sigma_c^+ \pi^-$, $\Sigma_c^0\pi^0$, 
 $\Sigma_c^0\rho^0$
 \\
$\Sigma_c^*=\Sigma_c(2520)$   &  $1$  &  $1(\frac32^+)$  &  2517 &
 $\Lambda_b \to$ $\Sigma_c^{*+} \pi^-$, $\Sigma_c^{*0}\pi^0$, 
 $\Sigma_c^{*0}\rho^0$
 \\
$\Xi_c$, $\Xi_c'$   &  $0$, $1$  &  $\frac12 (\frac12^+)$  &  2469, 2576 &
 $\Lambda_b \to$ $\Xi_c^{\prime0} K^0$
 \\
$\Xi_c^*=\Xi_c(2645)$   &  $1$  &  $\frac12 (\frac32^+)$  &  2646 &
 $\Lambda_b \to$ $\Xi_c^{*0} K^0$
 \\
\hline
 $\Theta_c$, $\Theta_b$  &  $1$  & $0(\frac12^+)$  &
   $m_{\Theta_{c}}$, $m_{\Theta_{b}}$  & 
 $\Theta_b^+ \to \Theta_c^0\pi^+$, $\Theta_c^0\to \Theta^+\pi^-$
 \\
 $\Theta_c^*$, $\Theta_b^*$  &  $1$  & $0(\frac32^+)$  & 
  $\sim m_{\Theta_{c}}+70$, & $\Theta_c^*\to \Theta_c\gamma$ \mbox{ or strongly}
 \\
 & & & $\sim m_{\Theta_{b}}+22$ &  $\Theta_b^*\to \Theta_b\gamma$   \\
\hline\hline
\end{tabular}
\caption{\setlength\baselineskip{18pt}
The decays considered in this paper.  Here $s_l$ is the spin of the light
degrees  of freedom \cite{Isgur:wq}. The mass shown for the $\Sigma_c^{(*)}$
is  the average of the charge 0 and $+1$ states, and for the $\Xi_c$'s the 
mass is the average in the doublet.  The stability of the pentaquark states 
$\Theta_Q(=\bar Q udud)$ and their value of $s_l$ are both conjectures.}
\label{tab:charm}
\end{table}

There is considerable experimental interest in these decays.  Recently the CDF
Collaboration measured~\cite{cdflambdab} 
\beq\label{cdfrate} 
{f_{\Lambda_b}\over f_d}\, {{\cal B}(\Lambda_b\to \Lambda_c^+\pi^-)\over 
  {\cal B}(\B0bar\to D^+\pi^-)} 
  = 0.66 \pm 0.11_{\rm (stat)} \pm 0.09_{\rm (syst)} \pm 0.18_{\rm (BR)}\,, 
\eeq 
where $f_{\Lambda_b}$ and $f_d$ are the fragmentation fractions of $b$ quarks
to $\Lambda_b$ and $\B0bar$, respectively.  Using the input $f_{baryon}/f_d =
0.304\pm 0.053$, CDF quoted ${\cal B}(\Lambda_b \to \Lambda_c^+\pi^-) / {\cal
B}(\B0bar \to D^+\pi^-) \simeq 2.2$~\cite{cdflambdab}. The measured lifetimes,
$\tau(\Lambda_b) = 1.23\,$ps and $\tau(B^0) = 1.54\,$ps~\cite{Hagiwara:fs},
then imply that $\Gamma(\Lambda_b \to \Lambda_c^+\pi^-) / \Gamma(\B0bar \to
D^+\pi^-) \simeq 2.7$. More experimental results on semileptonic and other
nonleptonic $\Lambda_b$ decays are expected in the near future.

The part of the weak Hamiltonian relevant for this paper is
\beq\label{Hweak}
H_W = {4G_F\over\sqrt2}\, V_{cb}V_{ud}^*\, \Big[ 
  C_1(m_b)\, O_1(m_b) + C_2(m_b)\, O_2(m_b) \Big] \,,
\eeq
where both the Wilson coefficients, $C_i$, and the four-quark operators 
\beq\label{fullops}
O_1 = (\bar c\, \gamma_\mu P_L b)\, (\bar d \gamma^\mu P_L u)\,, \qquad
O_2 = (\bar d\, \gamma_\mu P_L b)\, (\bar c \gamma^\mu P_L u)\,,
\eeq
depend on the renormalization scale which we take to be $m_b$, and $P_L =
(1-\gamma_5)/2$.  The combination $[C_1(m_b)+C_2(m_b)/3]\,|V_{ud}|$ is very
close to unity.

\begin{figure}[t]
\centerline{\includegraphics[width=.23\textwidth]{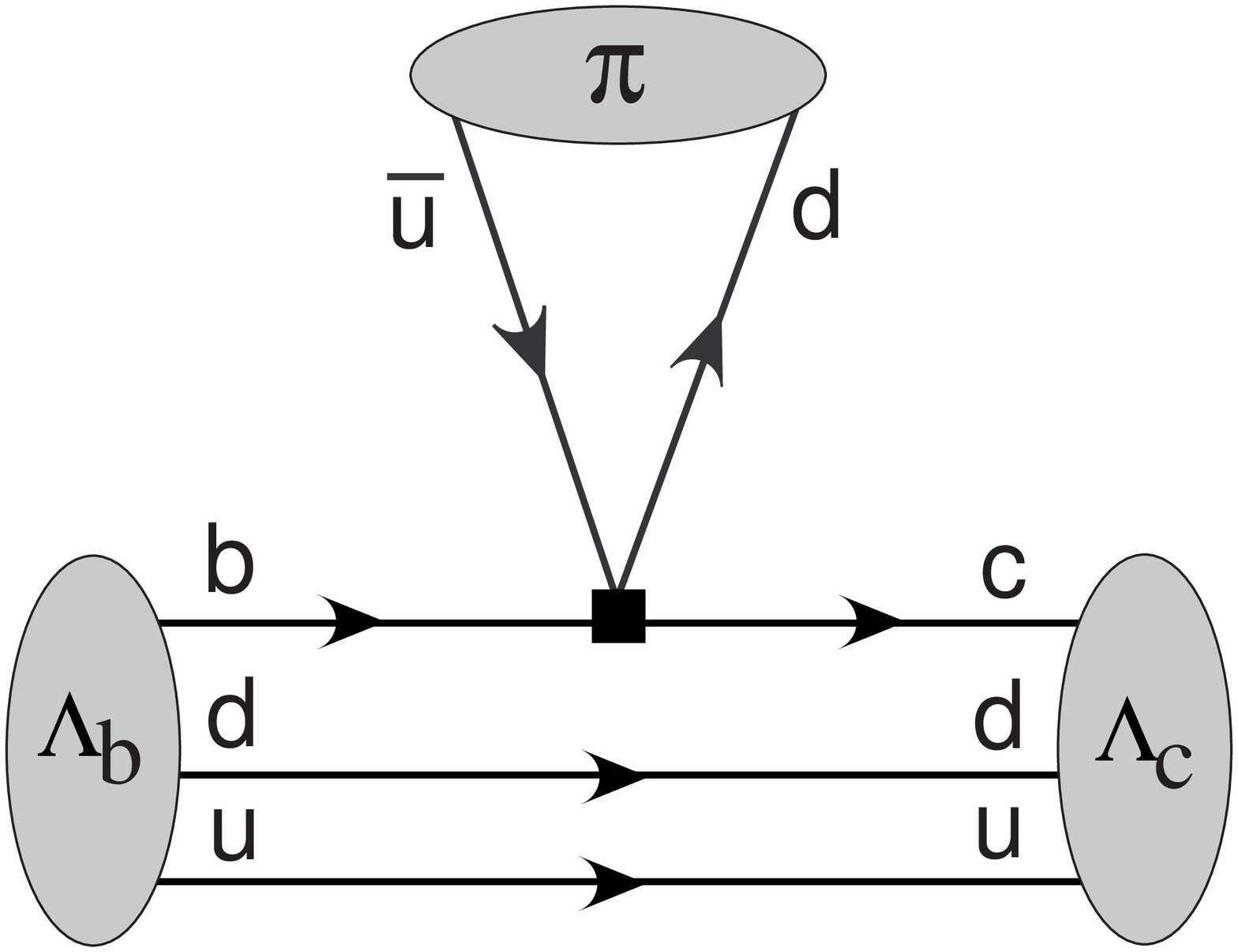} \hspace*{0.1cm}
  \includegraphics[width=.23\textwidth]{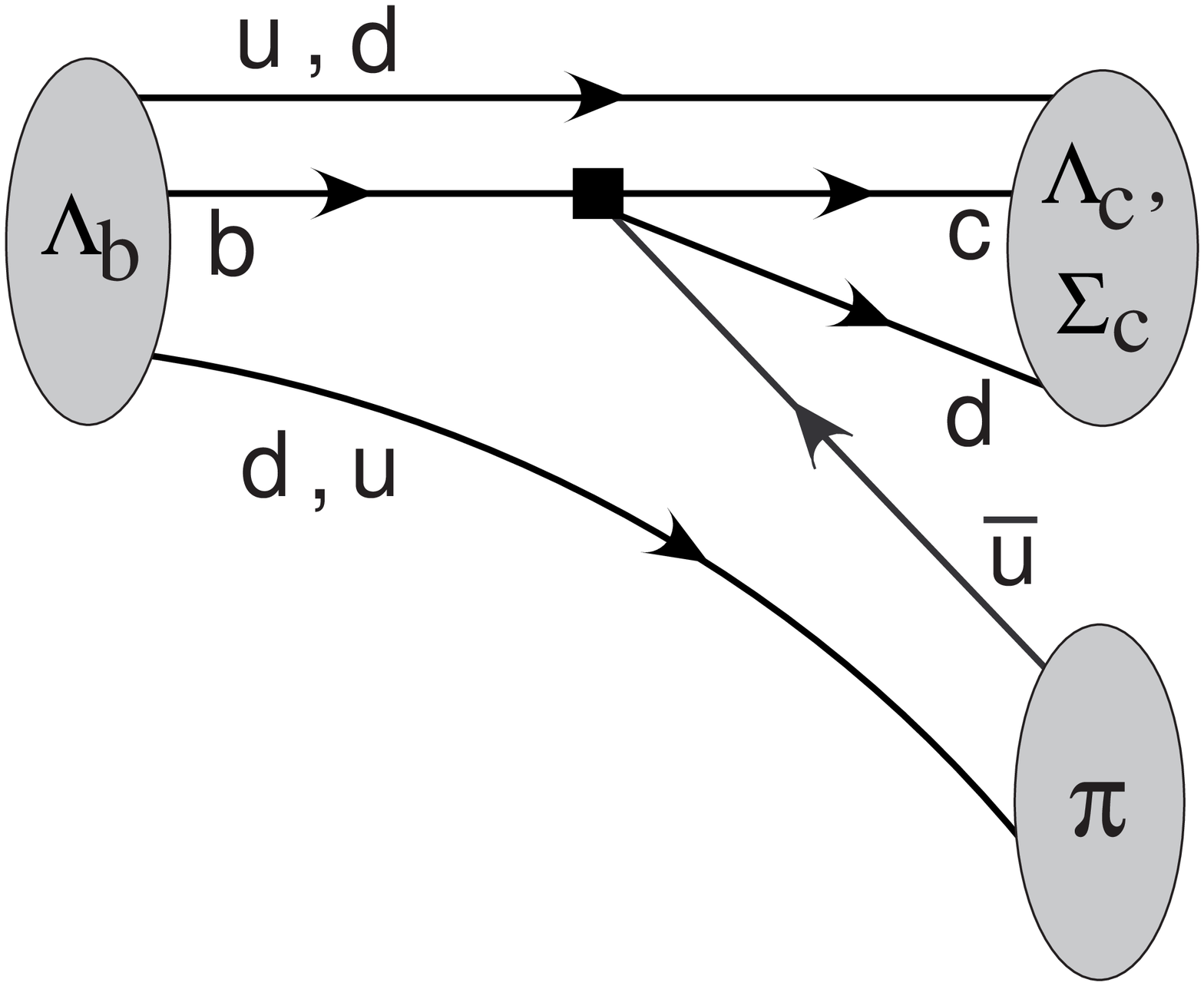} \hspace*{0.1cm}
  \includegraphics[width=.22\textwidth]{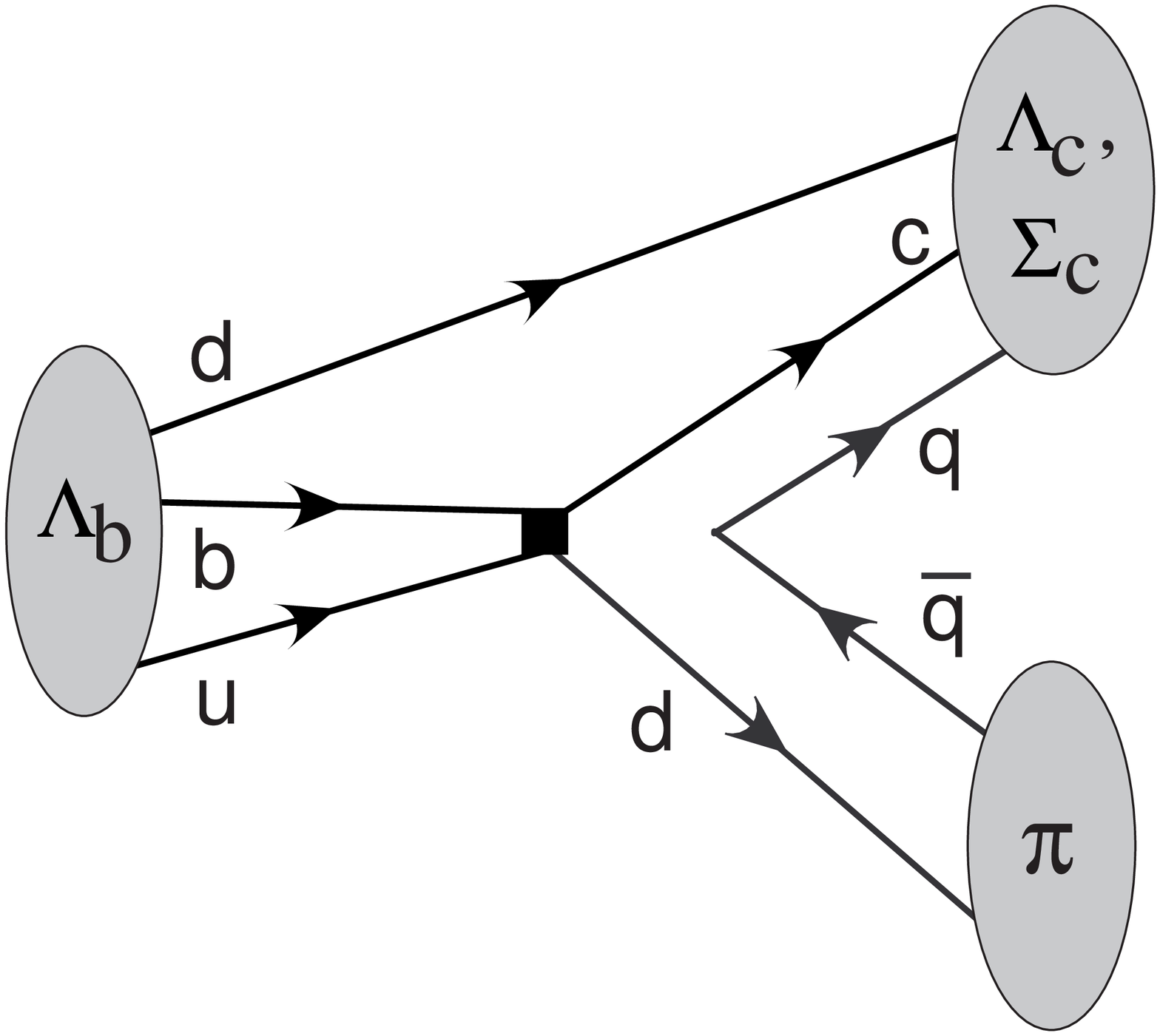} \hspace*{0.1cm}
  \includegraphics[width=.23\textwidth]{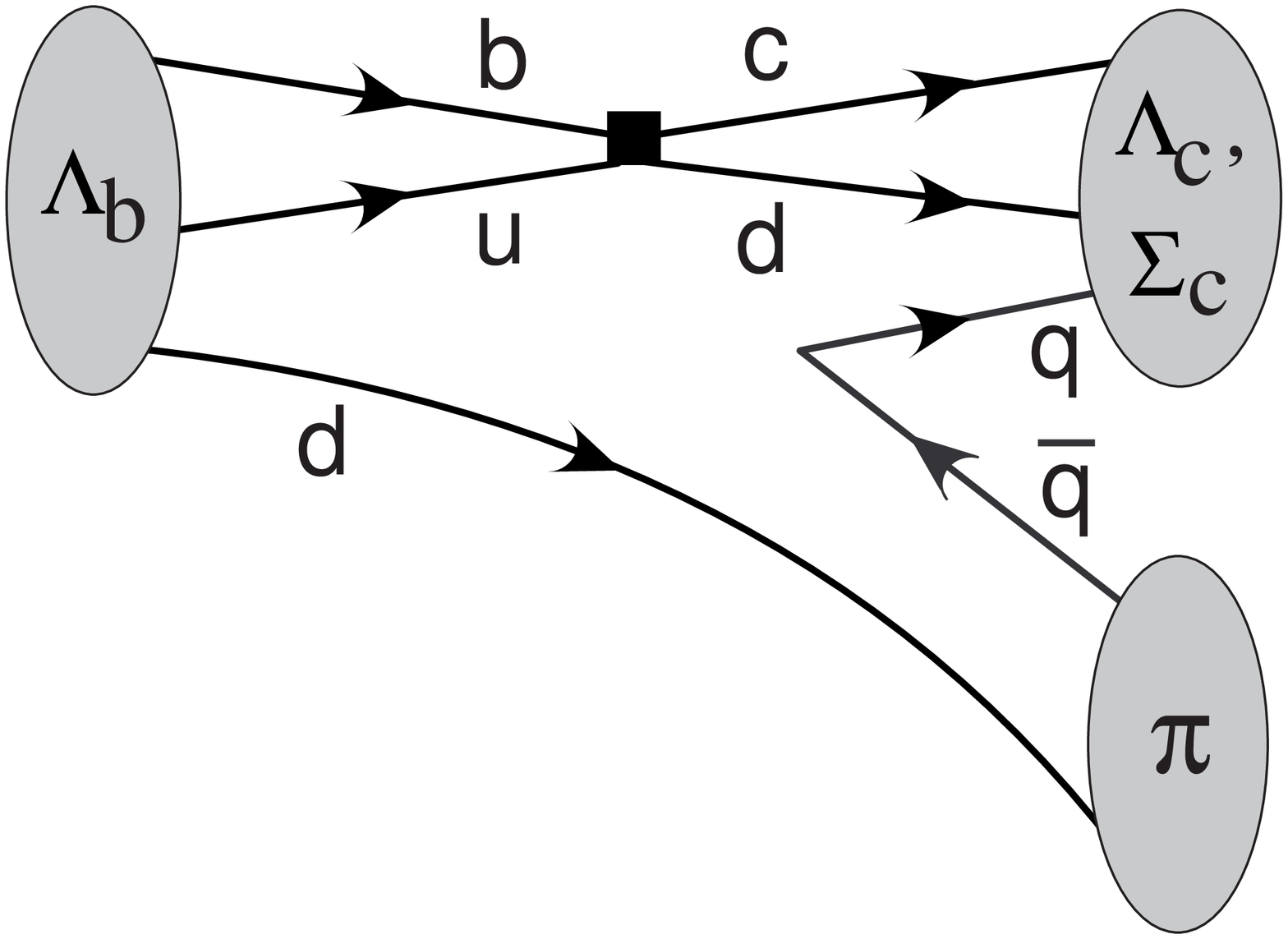}}
\vspace{-0.25cm}
\hbox{\hspace{1.6cm} Tree \hspace{2.0cm} Color-commensurate \hspace{1.1cm} 
  Exchange \hspace{2.25cm} Bow tie}
\caption{\setlength\baselineskip{18pt}
Diagrams for $\Lambda_b$ decays, giving amplitudes $T$, $C$, $E$, and $B$.
Decay to $\Lambda_c$ gets contributions from all four terms.  Decays to
$\Sigma_c^{(*)}$ and $\Xi_c$ do not have $T$ and $T,C$ contributions,
respectively.}
\label{fig:diagrams}
\end{figure}

Weak nonleptonic decays are sometimes characterized by diagrams corresponding
to different Wick contractions.  As shown in Fig.~\ref{fig:diagrams}, there are
more possibilities in baryon than in meson decays.  In particular, a ``Bow~tie"
contraction is unique to baryons.  The color structure for baryons also differs
from mesons: we find that the $C$ diagram is of the same order in the large
$N_c$ limit as the $T$ diagram.\footnote{If we treated the $N_c-3$ additional quarks 
in the baryons as flavors that are sterile under the weak interaction then 
color-commensurate would become color-suppressed.} Nonleptonic meson decay 
amplitudes are sometimes estimated
using naive factorization, which would set $\langle \Lambda_c\pi| O_1
|\Lambda_b\rangle = \langle \Lambda_c | \bar c \gamma_\mu P_L b | \Lambda_b
\rangle\, \langle \pi | \bar d \gamma_\mu P_L u |0\rangle$.  In baryon decays
the extra light quark implies that this procedure is ill-defined for all but
the tree diagram.  In naive factorization the $\Lambda_b\to \Sigma_c^{(*)} \pi$
decays are very suppressed, since the $T$ contribution vanishes separately in
the isospin and heavy quark limits~\cite{Politzer:ps} (just like the
semileptonic $\Lambda_b\to \Sigma_c^{(*)} \ell\bar\nu$ decays), the $C$
contribution vanishes after doing a Fiertz transformation on the four-quark
operator, and the $E$ and $B$ amplitudes are identically zero since the $u$ and
$b$ fields are in different quark bilinears.

In this letter we show that more rigorous techniques can still be applied to
make reasonable predictions for all these decays.  By expanding in $m_b, m_c,
E_\pi \gg \lqcd$ we show that for $\Lambda_b\to\Lambda_c^+\pi^-$  the
amplitudes corresponding to the diagrams in Fig.~\ref{fig:diagrams} satisfy $T
\gg C\sim E \gg B$, and we find that the experimental result in
Eq.~(\ref{cdfrate}) is consistent with theoretical expectations.  Next we
consider $\Lambda_b \to \Sigma_c^{(*)} \pi$ decays, and show the leading
contributions to these nonleptonic rates are suppressed by $\lqcd^2/E_\pi^2$,
much like in $B^0\to D^0\pi^0$.  Using heavy quark symmetry we derive a
relation between the decay rate to $\Sigma_c$ and $\Sigma_c^*$ and comment on
decays to $\Xi_c$. Finally we consider the detection of possible weakly
decaying heavy pentaquarks, $\Theta_b$ and $\Theta_c$, with nonleptonic decays.

The proof of factorization at leading order for $\Lambda_b \to \Lambda_c \pi$
decay follows closely that for $\B0bar \to D^{(*)+} \pi^-$~\cite{Bauer:2001cu},
so we do not review it here. In this case the nonperturbative expansion
parameter for SCET is $\lambda=\lqcd/E_{\pi}$~\cite{Dugan:1990de}. Since
$E_{\pi}$ is set by the bottom and charm quark masses, we take this to be of
the same order as the expansion parameter for the heavy quark effective theory
(HQET), i.e., $\lambda\sim \lqcd/m_Q$ ($Q=c,b$). Working at leading order in
$\lambda$ and $\alpha_s(m_b)$ and neglecting the pion mass, the $\Lambda_b \to
\Lambda_c \pi$ matrix element factorizes in the standard way, 
\beq\label{Lbfactor} 
\langle \Lambda_c(v',s')\, \pi | H_W | \Lambda_b (v,s)\rangle 
  = \sqrt2\, G_F \bigg(C_1 + \frac{C_2}{3}\bigg) V_{cb} V_{ud}^*\, 
  f_\pi E_\pi\, \langle \Lambda_c(v',s')|\, \bar c\, \nslash P_L b\,
  |\Lambda_b(v,s) \rangle \,, 
\eeq
where $f_{\pi}=131\,$MeV is the pion decay constant, $n$ is a light-like
four-vector along the direction of the pion's four-momentum, $p_{\pi}^{\mu} =
E_{\pi} n^\mu$, and the four-velocities of the $\Lambda_b$ and $\Lambda_c$ are
$v$ and $v'$, respectively.  Perturbative corrections induce a multiplicative
factor in Eq.~(\ref{Lbfactor}), $\langle T(x)\rangle_{\pi} = \int_0^1 \d x\,
T(x)\, \phi_\pi(x)$, where $T(x)$ is computable and $\phi_\pi$ is the
nonperturbative light-cone pion distribution
function~\cite{Politzer:au,Beneke:2000ry}, and a term proportional to the
matrix element of $\bar c\, \nslash P_R b$.  At leading order in
$\alpha_s(m_Q)$, we can set $\langle T(x)\rangle_{\pi}=1$ and the term
involving $\bar c\, \nslash P_R b$ to 0.  This implies that the nonleptonic
rate is related to the semileptonic differential decay rate at maximal recoil, 
\beq\label{Lbfactor2}
\Gamma(\Lambda_b\to \Lambda_c\pi) = {3\pi^2 (C_1+C_2/3)^2\, |V_{ud}|^2 f_\pi^2
  \over m_{\Lambda_b}^2\, r_\Lambda}\, \Bigg( {\d\Gamma(\Lambda_b\to
  \Lambda_c\ell\bar\nu)\over \d w} \Bigg)_{\wmax} \,, 
\eeq
where $r_\Lambda = m_{\Lambda_c}/m_{\Lambda_b}$, $w = v\cdot v' =
(m_{\Lambda_b}^2 + m_{\Lambda_c}^2 - q^2) / (2m_{\Lambda_b} m_{\Lambda_c})$,
and $\wmax$ corresponds to $q^2=m_\pi^2(\simeq 0)$.

The semileptonic $\Lambda_b\to \Lambda_c\ell\bar\nu$ form factors are
\beqa\label{Lbffdef}
\bra{\Lambda_c(p',s')} V_\mu \ket{\Lambda_b(p,s)} &=& \bar u(p',s') 
  \Big[ f_1 \gamma_\mu + f_2 v_\mu + f_3 v'_\mu \Big] u(p,s)\,, \nn\\
\bra{\Lambda_c(p',s')} A_\mu \ket{\Lambda_b(p,s)} &=& \bar u(p',s')
  \Big[ g_1 \gamma_\mu + g_2 v_\mu + g_3 v'_\mu \Big] \gamma_5\, u(p,s)\,,
\eeqa
where the $f_i$ and $g_i$ are functions of $w$, and the relevant currents are
$V_\nu = \bar c\gamma_\nu b$ and $A_\nu = \bar c\gamma_\nu\gamma_5 b$.  The
spinors are normalized to $\bar u(p,s)\gamma^\mu u(p,s) = 2p^\mu$.  In the
heavy quark limit,
\beqa
\zeta(w) &=& f_1(w) = g_1(w)\,, \nn\\
0 &=& f_2(w) = f_3(w) = g_2(w) = g_3(w)\,,
\eeqa
where $\zeta(w)$ is the Isgur-Wise function for ground state baryons.
The differential decay rate is given by
\beq\label{rate}
{\d\Gamma(\Lambda_b\to \Lambda_c\ell\bar\nu)\over \d w} = 
  {G_F^2\, m_{\Lambda_b}^5 |V_{cb}|^2\over 24\,\pi^3}\, 
  r_\Lambda^3\, \sqrt{w^2-1}\, \Big[6w + 6w r_\Lambda^2 
  - 4r_\Lambda - 8r_\Lambda w^2\Big]\, \Fb^2(w) \,,
\eeq
where in the $m_Q \gg \lqcd$ limit $\Fb(w)$ is equal to the Isgur-Wise
function, $\zeta(w)$, and in particular $\Fb(1) = 1$. In terms of the original
form factors
\begin{eqnarray}
\Fb(w)^2 &=& \Big[6w + 6w r_\Lambda^2 - 4r_\Lambda - 8r_\Lambda w^2\Big]^{-1}\,
  \bigg\{ (w-1)\, \Big[(1+r_\Lambda)f_1 + (w+1)(r_\Lambda f_2+f_3)\Big]^2\nn\\*
&&{} + (w+1)\, \Big[(1-r_\Lambda)g_1 - (w-1)(r_\Lambda g_2+g_3)\Big]^2\nn\\*
&&{} + 2(1-2r_\Lambda w+r_\Lambda^2) \Big[(w-1)f_1^2 + (w+1)g_1^2\Big] 
  \bigg\}\,.
\end{eqnarray}

Combining the above results for $\Lambda_b\to \Lambda_c^+ \pi^-$ decay with the
analogous ones for $\B0bar\to D^{(*)+}\pi^-$ we find that
\beq\label{factest}
{\Gamma(\Lambda_b \to \Lambda_c \pi^-) \over \Gamma(\B0bar \to D^{(*)+} \pi^-)}
= {8m_{\Lambda_b}^3 (1-r_\Lambda^2)^3\, r_{D^{(*)}} \over 
  m_B^3 (1-r_{D^{(*)}}^2)^3 (1+r_{D^{(*)}})^2} \left({\zeta(\wmax^{\Lambda}) 
  \over \xi(\wmax^{D^{(*)}})}\right)^2
\eeq
where $\xi$ is the Isgur-Wise function for $B \to D^{(*)}$ semileptonic decay,
and $r_{D^{(*)}} = m_{D^{(*)}}/m_B$.  
When the $\Lambda_b\to \Lambda_c^+ \ell^-
\bar\nu$ rate is measured, one can directly test factorization using
Eq.~(\ref{Lbfactor2}) or Eq.~(\ref{factest}).  In the absence of this data, we
have to resort to using model predictions for the baryon Isgur-Wise function. 
If the ratio of Isgur-Wise functions in Eq.~(\ref{factest}) is unity then the
prefactor in Eq.~(\ref{factest}) implies that $\Gamma(\Lambda_b \to \Lambda_c
\pi^-)/\Gamma(\B0bar \to D^{(*)+} \pi^-) = 1.6 (1.8)$.  This enhancement is in
rough agreement with the data in Eq.~(\ref{cdfrate}).  A similar result also
follows from the  small velocity limit ($m_Q \gg m_b-m_c \gg \lqcd$), in which
the nonleptonic rates satisfy $\Gamma(\Lambda_b\to \Lambda_c\pi) : \Gamma(B\to
D^*\pi) : \Gamma(B\to D \pi) = 2 : 1 : 1$, while for the semileptonic rates
$\Gamma(\Lambda_b \to \Lambda_c \ell\bar\nu) : \Gamma(B\to D^* \ell\bar\nu):
\Gamma(B\to D \ell\bar\nu) = 4 : 3 : 1$.

The large $N_c$ limit provides some support for the ratio of baryon to meson
Isgur-Wise functions being close to unity at maximal recoil. In the large $N_c$
limit the heavy baryons can be treated as bound states of chiral solitons and
mesons containing a heavy quark.  In this picture, the baryon Isgur-Wise
function, $\zeta(w)$, is predicted to be $\zeta(w) = 0.99\,
e^{-1.3(w-1)}$~\cite{Jenkins:1992se}.\footnote{Updating the parameters by
fitting the mass splitting to give $\kappa=(0.411\,{\rm GeV})^3$, and using
$m_N=\bar\Lambda=0.8\,{\rm GeV}$ (instead of $M_N$) for the mass of the light
degrees of freedom leaves the exponent essentially unchanged.}  This gives
$\zeta(\wmax^{\Lambda}=1.4) = 0.57$, which is indeed close to
$\xi(\wmax^{D^*}=1.5) \simeq 0.55$~\cite{Abe:2001cs}.  Using this model for
$\zeta(w)$, $|V_{cb}| = 0.04$, $\tau_{\Lambda_b} = 1.23\,$ps, and
Eqs.~(\ref{Lbfactor2}) and (\ref{rate}) yield the prediction that ${\cal B}
(\Lambda_b\to \Lambda_c^+ \pi^-) = 4.6 \times 10^{-3}$.  As expected, this is
larger than ${\cal B}(\B0bar\to D^{(*)+}\pi^-) \simeq 2.7 \times 10^{-3}$.
However, the uncertainty in this prediction is quite large, particularly given
that large $N_c$ strictly only applies for $w$ near 1. The same large $N_c$
inputs predict ${\cal B} (\Lambda_b\to \Lambda_c^+ \ell^- \bar\nu) \approx
6\%$, i.e., this channel is expected to make up a large part of the inclusive
$\Lambda_b\to X_c \ell^- \bar\nu$ rate, with the $s_l^P=1^-$ excited
$\Lambda_c$ states making up a significant fraction of the
remainder~\cite{Leibovich:1997az}.

Order $\lqcd/m_Q$ corrections to these predictions may be significant.  The
$\Lambda_b\to \Lambda_c^+ \pi^-$ amplitude receives contributions from the $T$,
$C$, $E$, and $B$ classes of diagrams in Fig.~\ref{fig:diagrams}.  In SCET,
$|E/T|$ and $|C/T|$ are of order $\lqcd/m_Q$~\cite{Mantry:2003uz}, and we will
show later that $|B/T|$ is further suppressed.  In $B\to D\pi$ decay we know
from ${\cal B}(B^-\to D^0\pi^-) / {\cal B}(\B0bar\to D^+\pi^-) \simeq
1.8$~\cite{Hagiwara:fs} that $\lqcd/m_Q$ corrections affect the amplitudes at
the $15-30\%$ level. In particular $| A(\bar B^0\to D^+\pi^-)| = |T+E| = (5.9
\pm 0.3)\times 10^{-7}\,{\rm GeV}$ and $| A(B^-\to D^0\pi^-)| = |T+C| = (7.7
\pm 0.3)\times 10^{-7}\,{\rm GeV}$.  The ratio of these amplitudes can be
reproduced by a power correction involving a hadronic parameter $|s_{\rm eff}|
\simeq 430\,{\rm MeV}$, which is of natural size~\cite{Mantry:2003uz}.  Since
$B_s\to D_s^-\pi^+$ only has a $T$ contribution, accurate measurement of this
rate will improve our understanding of the size of $E$ and $C$.  CDF recently
measured $[f_s\, {\cal B}(B_s\to D_s^-\pi^+)] / [f_d\, {\cal B}(B^0\to
D^-\pi^+)] = 0.35 \pm 0.05_{\rm (stat)} \pm 0.04_{\rm (syst)} \pm 0.09_{\rm
(BR)}$~\cite{cdfds}, and using $f_s/f_d = 0.26\pm 0.03$ yields ${\cal B}(B_s\to
D_s^-\pi^+)/ {\cal B}(B^0\to D^-\pi^+) \simeq 1.35$. Neglecting $SU(3)$
breaking\footnote{In the heavy quark limit of the $T$ amplitude $SU(3)$ will be
tested by the measurement of $B_s\to D_s \ell\bar\nu$.} this implies $|A(B_s\to
D_s^-\pi^+)|=|T|=(7.3\pm 1.5)\times 10^{-7}\,{\rm GeV}$ and that $|C|$ and
$|E|$ may be comparable. The errors are still too large to draw any definite
conclusions.

Now we turn to $\Lambda_b \to \Sigma_c\pi$ decays. As shown in
Table~\ref{tab:charm}, there are two $\Sigma_c$ states with different spin
which we refer to as $\Sigma_c$ and $\Sigma_c^*$.  They form a heavy quark spin
symmetry doublet with the spin and parity of the light degrees of freedom,
$s_l^{\pi_l} = 1^+$.  Under isospin, the $\Lambda_b$ is $I = 0$, the
$\Sigma_c^{(*)}$ is $I = 1$, and the Hamiltonian is $I=1$, so the
$\Sigma_c^{(*)}\pi$ final state must be $I=1$ (it can not be $I=0$ or 2).
Therefore the rates to the two different charge channels are equal, 
\beq
  \Gamma(\Lambda_b \to \Sigma_c^{(*)0} \pi^0) 
  = \Gamma(\Lambda_b \to \Sigma_c^{(*)+} \pi^-) \,.  
\eeq 
Based on $B$ decay data and the SCET power counting, we expect
$\Gamma(\Lambda_b \to \Sigma_c^{(*)}\pi)$ to be up to about an order of
magnitude smaller than $\Gamma(\Lambda_b\to\Lambda_c\pi)$, since the leading
contributions to $\Lambda_b\to\Sigma_c^{(*)}\pi$ are power suppressed.

\begin{figure}[t]
\centerline{\includegraphics[width=.35\textwidth]{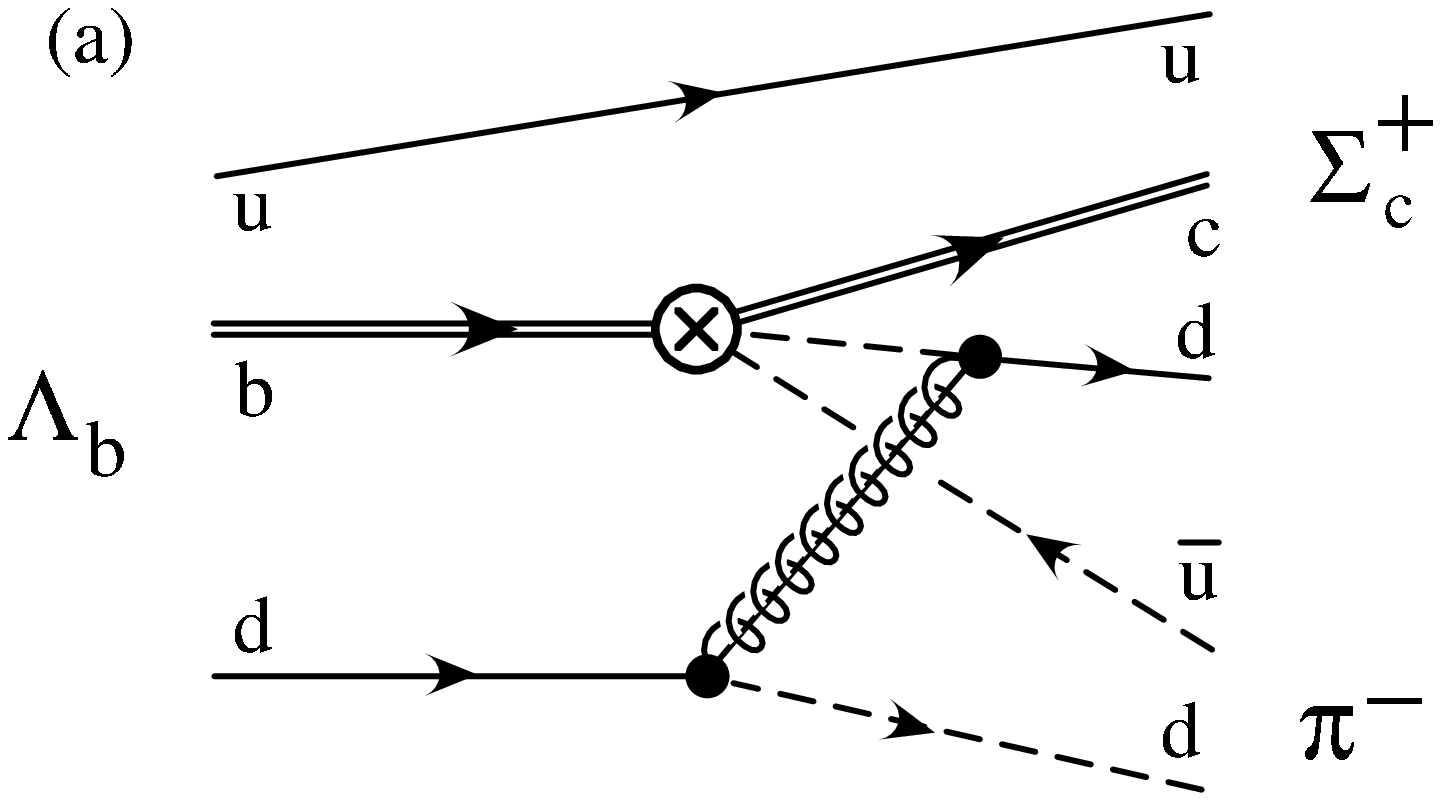} \hspace*{1cm}
  \includegraphics[width=.35\textwidth]{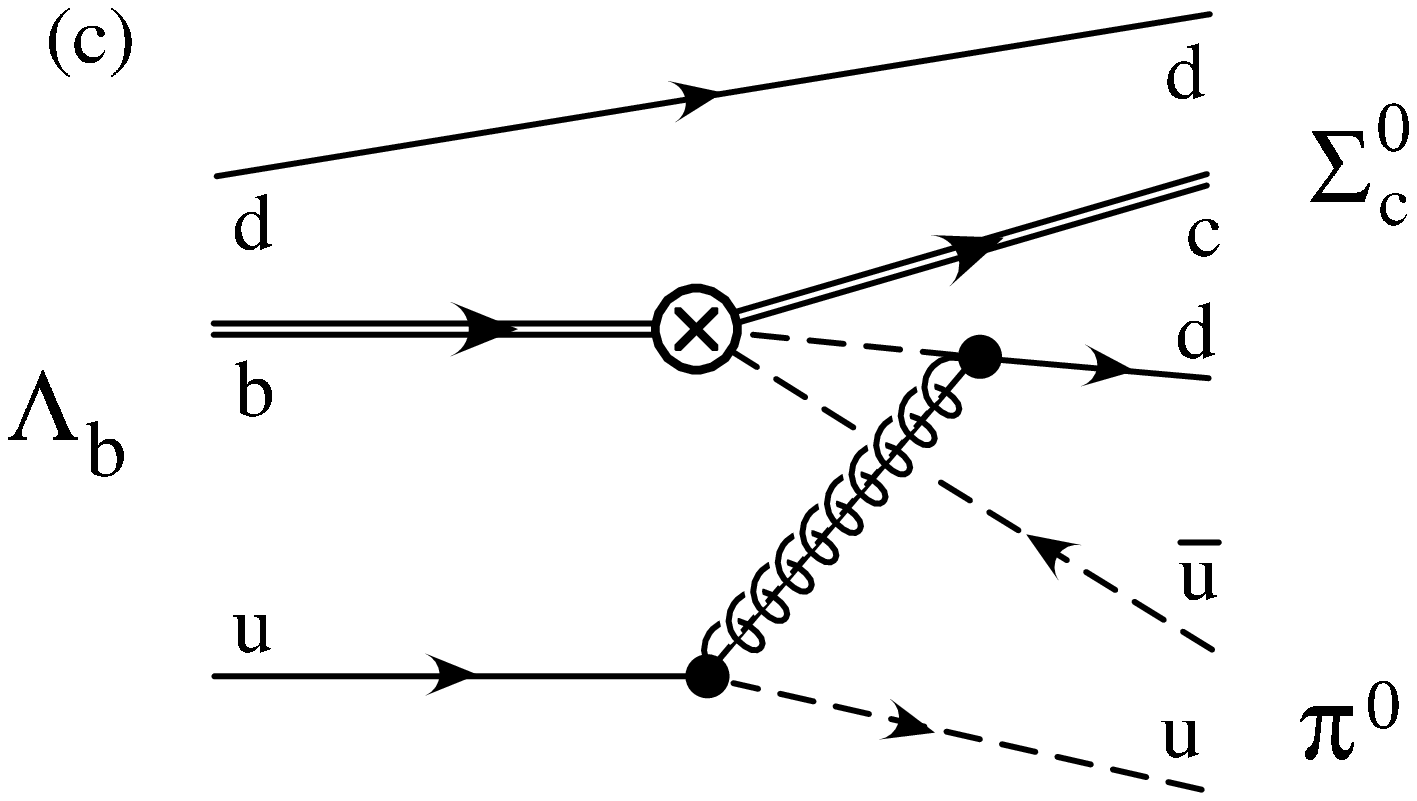}}
\vspace*{.25cm}
\centerline{\includegraphics[width=.35\textwidth]{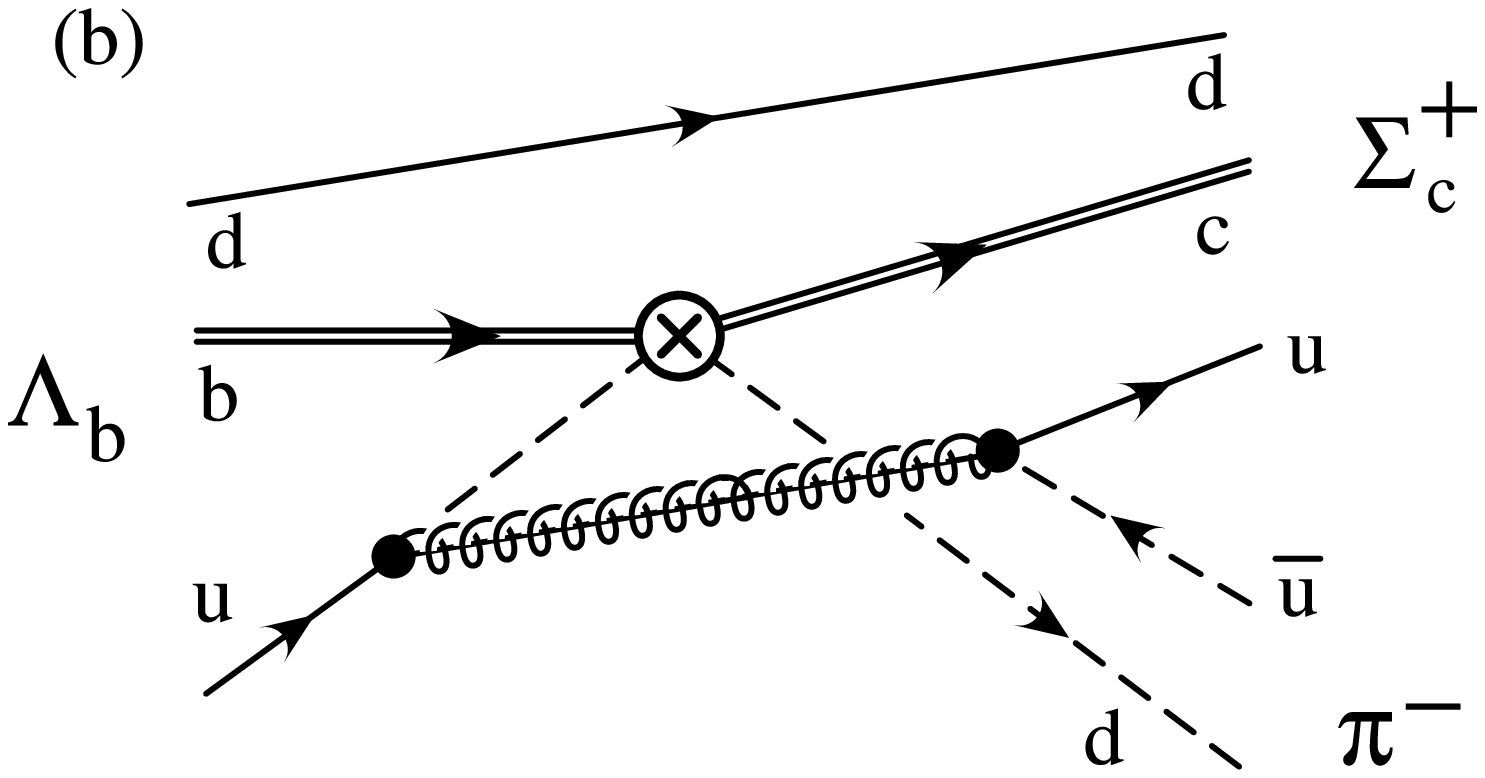} \hspace*{1cm}
  \includegraphics[width=.35\textwidth]{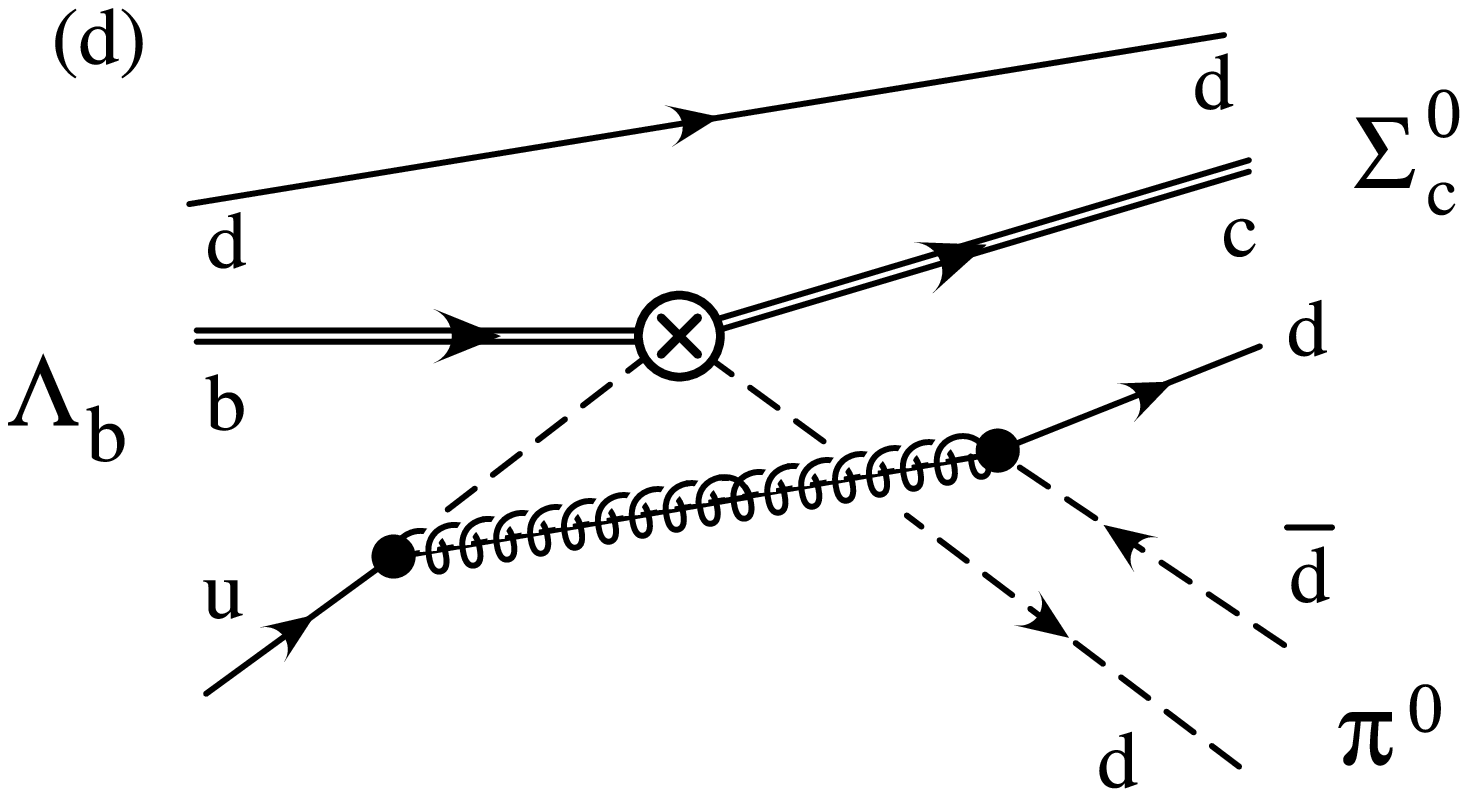}}
\caption{\setlength\baselineskip{18pt}
Contributions in \SCETa\ to $\Lambda_b\to \Sigma_c^+\pi^-$ [(a) and (b)], and
to $\Sigma_c^0\pi^0$ [(c) and (d)].  Solid dots denote insertions of the
suppressed usoft-collinear Lagrangian, ${\cal L}_{\xi
q}^{(1)}$, the double lines are heavy quarks, the dashed
lines are collinear quarks, the solid lines are usoft quarks, and the ``looped
lines" are collinear gluons. The nonleptonic weak vertex is denoted by
{\boldmath $\otimes$}.}
\label{fig:Sigmafact}
\end{figure}

Again, we use SCET to expand in $\lqcd/m_Q$, $\lqcd/E_\pi$, and
$\alpha_s(m_Q)$, keeping only the leading terms that cause the $\Lambda_b \to
\Sigma_c^{(*)}\pi$ transitions.  These come from the $C$ and $E$ diagrams in
Fig.~\ref{fig:diagrams} and their contributions can be studied following the
analysis of $\bar B^0\to D^{(*)0}\pi^0$ in Ref.~\cite{Mantry:2003uz}.  The
leading diagrams in \SCETa\ that determine the matching onto power suppressed
operators are shown in Fig.~\ref{fig:Sigmafact}.  To match the $C$ and $E$
diagrams, two insertions of the mixed usoft-collinear Lagrangian, ${\cal
L}_{\xi q}^{(1)}$~\cite{Beneke:2002ph}, is required, each yielding a
suppression of $\sqrt{\lqcd/E_\pi}$.  This yields the power counting that
$|C/T|$ and $|E/T|$ are ${\cal O}(\lqcd/E_\pi)$.  In contrast, matching the $B$
diagram in Fig.~\ref{fig:diagrams} requires four insertions of ${\cal L}_{\xi
q}^{(1)}$ (or other higher dimensional terms in the Lagrangian), and $B$ is
therefore power suppressed compared to $C$ and $E$ by at least an additional
$\lqcd/E_\pi$.

In addition there is a further matching onto \SCETb. The resulting matrix
element involves soft and collinear operators which
factor~\cite{Mantry:2003uz}.\footnote{Since the messenger modes from
Ref.~\cite{BHLN} do not spoil factorization for cases with a product of color
singlet soft and collinear operators, we can neglect them in our analysis.} The
matrix element of the weak Hamiltonian, $\langle \Sigma_c^0(v',s') \pi^0\, |
H_W |\Lambda_b (v,s)\rangle$, can be written (neglecting $\alpha_s(m_Q)$
corrections) as a convolution integral of a jet function, $J(x,k_1^+,k_2^+)$,
with the matrix element involving the collinear fields, $\langle \pi | O_c(x) |
0\rangle$ which gives $\phi_\pi(x)$, and that involving the soft fields,
$\langle \Sigma_c(v',s') | O_s(k_j^+) | \Lambda_b (v,s)\rangle$.  In what
follows we only need the form of the soft operator~\cite{Mantry:2003uz}
\beq\label{Os} 
O_s(k_j^+) = \Big[(\bar h_{v'}^{(c)} S) \nslash P_L (S^\dagger h_v^{(b)})\Big] 
  \Big[(\bar d S)_{k_1^+} \nslash P_L (S^\dagger u)_{k_2^+}\Big] \,, 
\eeq
where $h_v^{(Q)}$ is an HQET heavy quark field, $S$ is a soft Wilson line, and
the subscripts denote the momentum carried by the fields.  For our purposes the
most important aspect of the analysis is that $O_s$ only involves the the
combination $\bar h_{v'}^{(c)}\, \nslash P_L\, h_v^{(b)}$. Thus, by heavy quark
symmetry
\begin{eqnarray}\label{Sigcrate}
\langle \Sigma_c(v',s') | O_s |\Lambda_b (v,s)\rangle
&=& {1 \over \sqrt 3}\, \bar u_{\Sigma_c}(v',s')\, (\gamma^\mu-v^{\prime\mu})\,
  \gamma_5 \nslash P_L\, u_{\Lambda_b}(v,s)\, X_\mu,\nn\\*
\langle \Sigma_c^*(v',s') | O_s |\Lambda_b (v,s)\rangle
&=& \bar u_{\Sigma_c^*}^\mu(v',s')\, \nslash P_L\, u_{\Lambda_b}(v,s)\, X_\mu,
\end{eqnarray}
where $v$ and $v'$ are the four-velocities of the $\Lambda_b$ and
$\Sigma_c^{(*)}$ respectively. The spinor field normalizations are $\bar
u(v,s)\, u(v,s)=1$ for the $\Lambda_b$ and $\Sigma_c$, and $\bar
u_\alpha(v,s)\, u^\alpha(v,s)=-1$ for the $\Sigma_c^*$. $X_\mu$ is the
most general vector compatible with the symmetries of QCD,
\beq\label{Xdef}
  X_\mu = a\, n_\mu + b\, v_\mu + c\, v'_\mu\,.
\eeq
Note that in Eq.~(\ref{Os}) the part of $O_s$ involving the light quark fields
is parity violating, so we need not worry about the fact that $\Lambda_b\to
\Sigma_c$ is an ``unnatural" transition.  Using $m_{\Lambda_b} v = m_{\Sigma_c}
v' + E_\pi n$ to eliminate the term proportional to $v_\mu$ in
Eq.~(\ref{Xdef}), it is easy to see that any term in $X_\mu$ proportional to
$v'_\mu$ does not contribute, so only $n_\mu$ remains.  Hence the ratio of the
rates for $\Lambda_b\to \Sigma_c\pi$ and $\Lambda_b\to \Sigma_c^*\pi$ are
determined model independently at leading order in $\lqcd/m_Q$ and
$\alpha_s(m_Q)$, similar to the $\B0bar \to D^{(*)0}\pi^0$ case.  We find
\beq\label{tworatio}
  {\Gamma(\Lambda_b\to\Sigma_c^*\pi) \over
  \Gamma(\Lambda_b\to\Sigma_c\pi) } 
  = 2 + {\cal O} \Big[\lqcd/m_Q\,,\, \alpha_s(m_Q) \Big] .
\eeq
To evaluate the square of the matrix element in Eq.~(\ref{Sigcrate}), we
used the spin sums from Ref.~\cite{Leibovich:1997az} for the various spin 
$\Sigma_c^{(*)}$ states.  The explicit calculation shows that the rate to
$\Sigma_c^*$ with $|s'|=3/2$ vanishes, as required by angular momentum
conservation.

A practical complication in testing this prediction is that the
$\Sigma_c^{(*)}$ states decay to $\Lambda_c \pi$, and so both decay channels
$\Lambda_b \to \Sigma_c^{(*)0} \pi^0 \to \Lambda_c \pi^-\pi^0$ and $\Lambda_b
\to \Sigma_c^{(*)+} \pi^- \to \Lambda_c \pi^0\pi^-$ contain a $\pi^0$ that
makes the reconstruction hard at hadron colliders.  This can be circumvented by
studying $\Lambda_b \to \Sigma_c^{(*)0} \rho^0$ decays. In this case the final
states are $\Lambda_c \pi^-\pi^+\pi^-$. Decays to a vector meson are
potentially more complicated due to the fact that ``long-distance''
contributions can induce transverse polarizations at the same order in
$\lqcd/E_{\pi}$.  However, at leading order in $\alpha_s(m_Q)$ these
long-distance contributions vanish for the $\rho^0$ final
state~\cite{Mantry:2003uz} and we obtain
\beq\label{tworho}
{ \Gamma(\Lambda_b\to\Sigma_c^{*0}\rho^0) \over
  \Gamma(\Lambda_b\to\Sigma_c^0\rho^0) } = 2 + {\cal O} \Big[\lqcd/m_Q\,,\,
\alpha_s(m_Q) \Big] .  
\eeq
It is also worth noting that
\beq
{\Gamma(\B0bar \to D^0 \pi^0) \over \Gamma(\B0bar \to D^0 \rho^0)} =
{\Gamma(\Lambda_b\to\Sigma_c^0\pi^0) \over \Gamma(\Lambda_b\to\Sigma_c^0\rho^0)
} + {\cal O} \Big[\lqcd/m_Q\,,\, \alpha_s\Big(\sqrt{m_Q\lqcd}\,\Big) \Big]\,,
\eeq
where in contrast to Eqs.~(\ref{tworatio}) and (\ref{tworho}) this prediction
requires a perturbative expansion at the intermediate scale $\sqrt{\lqcd\,
m_Q}$.

The decays $\Lambda_b\to \Xi_c K$ decays are also Cabibbo-allowed.  (These
decays involve ``$\bar s s$ popping" so only $\Xi_c^0 K^0$ is allowed, not
$\Xi_c^+ K^-$).  They are similar to $\Lambda_b\to \Sigma_c^{(*)}\pi$ in the
sense that the leading contribution in the heavy quark limit vanishes.   As
shown in Table~\ref{tab:charm} there are three $\Xi_c$ ``ground states",
$\Xi_c$, $\Xi_c'$, and $\Xi_c^*$.  The $\Xi_c$ and $\Xi'_c$ can mix, but the
former is expected to be mostly the state that transforms as {\boldmath
$\overline{3}$} under flavor $SU(3)$, while the latter is mostly a {\boldmath
$6$}.  The $\Xi^*_c$ also transforms as a {\boldmath $6$}, and forms a heavy
quark spin symmetry doublet with the $\Xi'_c$.  Thus, a relation similar to
Eq.~(\ref{tworatio}) also holds in this case, i.e., $\Gamma(\Lambda_b \to
\Xi^*_c K) / \Gamma(\Lambda_b \to \Xi'_c K) = 2 + {\cal O}[\lqcd/m_Q,\,
\alpha_s(m_Q)]$.  This prediction may be hard to test since $\Xi'_c$ decays to
$\Xi_c\gamma$.  One can also consider Cabibbo-suppressed $\Lambda_b$ decays,
e.g., $\Lambda_b \to \Xi_c\pi$, and the weak decays of other baryons containing
a heavy bottom quark.

Perhaps the most exciting possibility is the existence of heavy baryonic
pentaquark states. Recently several experiments claimed to observe a baryon
$\Theta^+(1540)$ with the quantum numbers of $K^+n$.  A possible explanation is
to consider the $\Theta^+$ as a bound state of two spin-zero $ud$ diquarks in a
P-wave with an $\bar s$ antiquark~\cite{Jaffe:2003sg}.  If diquarks play an
important role in these exotic states then the analogous heavy flavor states,
$\Theta_c = \bar c\, [ud]^2$ and $\Theta_b = \bar b\, [ud]^2$, may be below
threshold for strong decays by $\Delta E \simeq -100\,{\rm MeV}$ and $\Delta E
\simeq -160\,{\rm MeV}$ respectively~\cite{Jaffe:2003sg}.\footnote{It is possible
that the $\Theta_Q$ are above the strong decay
thresholds~\cite{Karliner:2003si}.  The assumptions in our analysis are that
(i) $\Theta_Q$ decay weakly; and (ii) the spin of the light degrees of freedom
is $s_l=1$, as suggested by~\cite{Jaffe:2003sg}.  If (i) is correct but (ii) is
not, it would be easy to modify our predictions, including
Eq.~(\ref{factest2}).}  Since the spin of the light degrees of freedom is
$s_l=1$, we expect from heavy quark symmetry that $\Theta_Q$ come with a
doublet partner of similar mass, $\Theta_Q^*$, as shown in
Table~\ref{tab:charm}, with a mass splitting of order $\lqcd^2/m_Q$.  From the
mass splittings for the $\Sigma_c$ and $\Xi_c$ we expect $m_{\Theta_c^*} -
m_{\Theta_c}\sim 70\,$MeV and $m_{\Theta_b^*} - m_{\Theta_b}\sim 22\,$MeV.  In
this case the $\Theta_Q^*$ may also be stable with respect to the strong
interactions and decay to $\Theta_Q\gamma$.  Since the splitting for
$\Theta_c^{(*)}$ is larger, it is possible that the $\Theta_c^*$ is just above
the strong decay threshold, making the spectroscopy even more interesting (like
in $D^*$ decays).

\begin{figure}[t]
\centerline{\includegraphics[width=.4\textwidth]{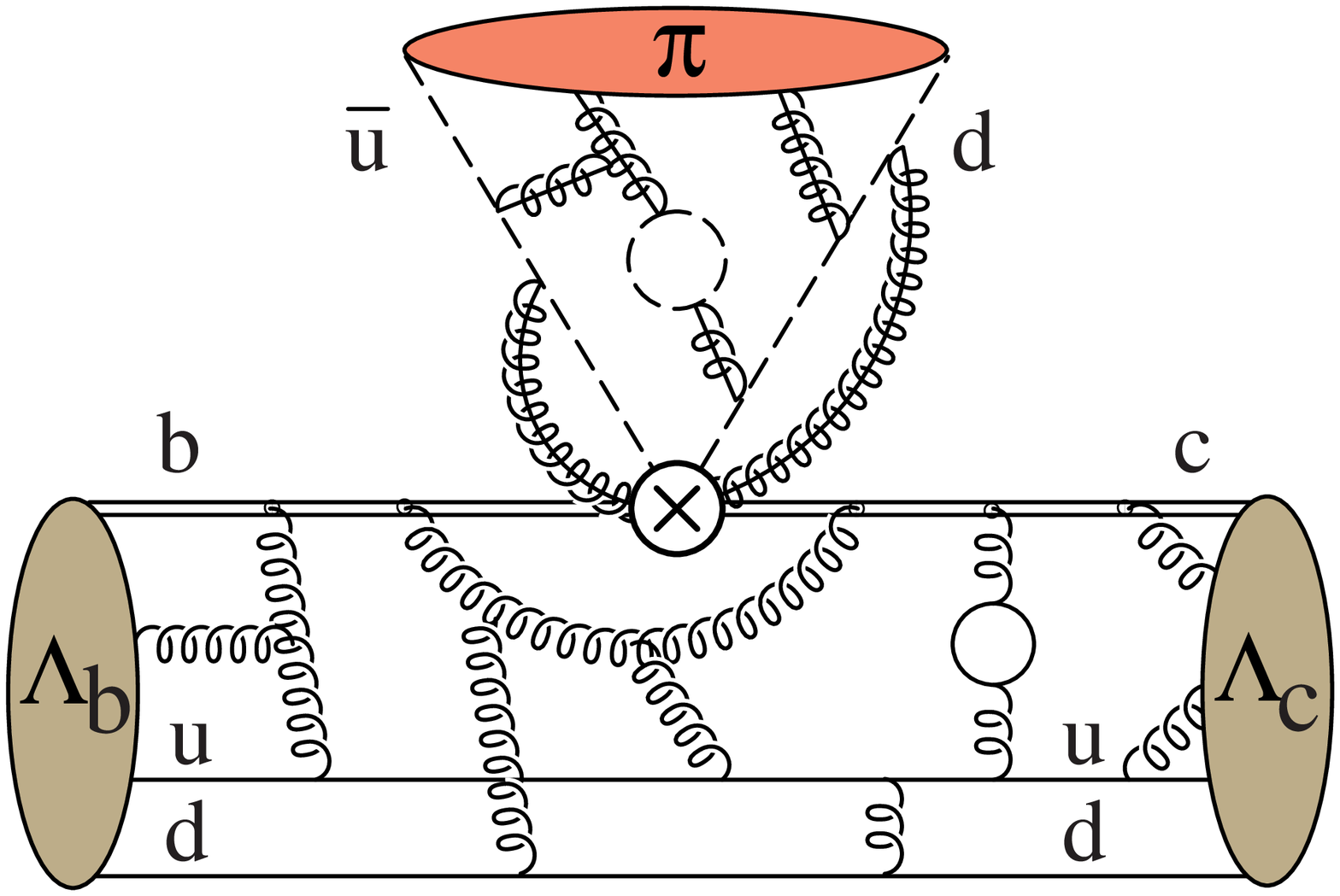} \hspace*{1.5cm}
  \includegraphics[width=.35\textwidth]{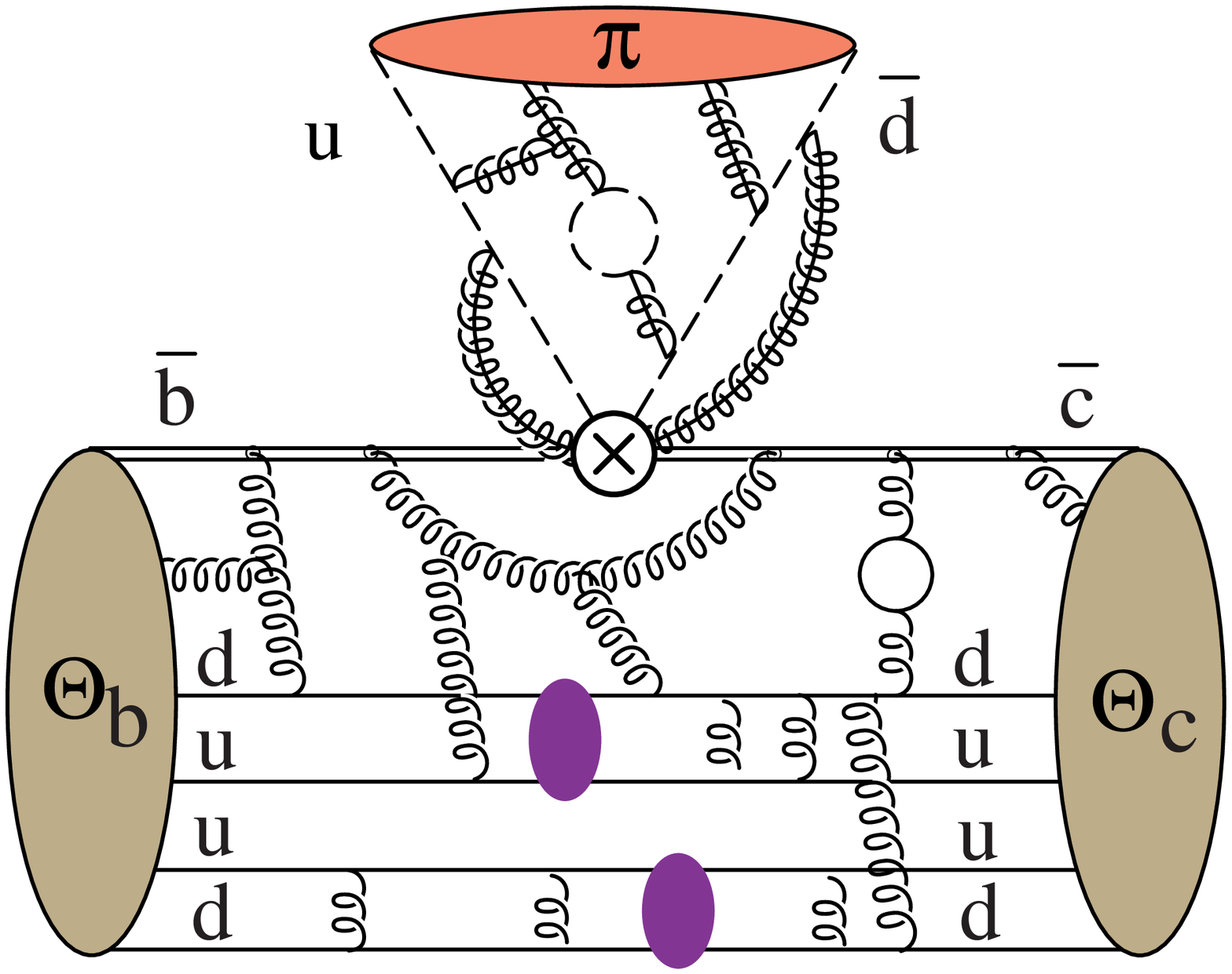} \hspace*{-0.5cm}} 
\caption{Comparison of the weak nonleptonic decays of the $\Lambda_b$ and
$\Theta_b$.}
\label{fig:penta}
\end{figure}

It may be possible to discover the $\Theta_{b,c}$ via the decay chains
\beq\label{Thetadecay} 
\Theta_b^+ \to \Theta_c^0 \pi^+\, \qquad \mbox{and}\qquad
\Theta_c^0 \to \Theta^+ \pi^- \to K_S\, p\, \pi^- \to \pi^+ \pi^- p\, \pi^- 
\eeq
that are Cabibbo-allowed and lead to all charged final states. The most
interesting aspect of the $\Theta_b^+ \to \Theta_c^0$ decay is that in the
diquark picture the correlation is maintained, as shown in
Fig.~\ref{fig:penta}, and so no additional suppression factor is expected.  In
weak $\Theta_b$ decays to ordinary baryons this would not be the case.  While
we do not know the $\Theta_Q$ production rates, we can estimate the branching
ratios in Eq.~(\ref{Thetadecay}). The lifetime of a weakly decaying
$\Theta_{b,c}$ is expected to be comparable with other weakly decaying hadrons
that contain a charm or a bottom quark.  The $\Theta_b^+ \to \Theta_c^0 \pi^+$
amplitude factorizes, and is related to $\Theta_b^+ \to \Theta_c^0 \ell\bar\nu$
via a formula identical to Eq.~(\ref{Lbfactor2}).  For the nonleptonic rate we
obtain in the heavy quark limit,
\begin{eqnarray}\label{factest2} 
{\Gamma(\Theta_b^+ \to
  \Theta_c^0 \pi^+) \over \Gamma(\Lambda_b \to \Lambda_c \pi^-)} &=&
{m_{\Theta_b}^3 (1-r_\Theta^2)^3\, \over m_{\Lambda_b}^3 (1-r_\Lambda^2)^3}\,
{1\over \zeta(\wmax^{\Lambda})^2}\, {1\over 144 r_\Theta^4}\, \bigg\{ 4
\Big[\eta_1(\wmax^\Theta)\Big]^2\, r_\Theta^2
(1+18r_\Theta^2+r_\Theta^4) \\
&&{} - 4 \eta_1(\wmax^\Theta)\eta_2(\wmax^\Theta)\,
r_\Theta(1-r_\Theta^2)^2(1+r_\Theta^2) + \Big[\eta_2(\wmax^\Theta)\Big]^2\,
(1-r_\Theta^2)^4 \bigg\}\,,\nn 
\end{eqnarray} 
where $r_\Theta = m_{\Theta_c}/m_{\Theta_b}$, and $\eta_{1,2}$ are the two
Isgur-Wise functions that parameterize the weak $\Theta_b \to \Theta_c^{(*)}$
matrix elements where $\eta_1(1) = 1$. In particular
\beq
 \langle \bar\Theta_c(v',s') |\, \bar c \Gamma b\, |\bar\Theta_b
(v,s)\rangle = {1 \over 3} \Big[ g^{\alpha\beta} \eta_1(w) - v^\alpha
v^{\prime\beta} \eta_2(w) \Big] \bar u(v',s')\, \gamma_5
(\gamma_\alpha+v'_\alpha) \Gamma (\gamma_\beta+v_\beta) \gamma_5 \, u(v,s) \,.
\eeq
Thus, ${\cal B}(\Theta_b^+ \to \Theta_c^0 \pi^+)$ is expected to be similar to
${\cal B}(\Lambda_b \to \Lambda_c \pi)$.  If the $\Theta_Q$ states exist then
an analysis of the $\lqcd/m_Q$ corrections would be warranted, as the mass of
the light degrees of freedom is sizable.  We expect ${\cal B}(\Theta_c^0 \to
\Theta^+ \pi^-)$ to be at the few percent level, while the other branching
ratios in Eq.~(\ref{Thetadecay}) may be of order unity.

In summary, we studied nonleptonic $\Lambda_b$ decays to $\Lambda_c\pi$,
$\Sigma_c\pi$ and $\Sigma_c^*\pi$.  Eqs.~(\ref{factest}), (\ref{tworatio}),
(\ref{tworho}), and (\ref{factest2}) are our main results.  In the $m_Q \gg
\lqcd$ limit the $\Lambda_b \to \Lambda_c\pi$ rate is related to $\Lambda_b \to
\Lambda_c\ell\bar\nu$, and we found that $\Gamma(\Lambda_b\to\Lambda_c\pi)$ is
expected to be larger than $\Gamma(B\to D^{(*)}\pi)$, as observed by CDF. At
leading order in $\lqcd/m_Q$ the $\Lambda_b \to \Sigma_c^{(*)}\pi$ rates
vanish, but an analysis of the leading contributions suppressed by $\lqcd/m_Q$
was still possible.  We predict $\Gamma(\Lambda_b \to \Sigma_c^*\pi) \,/\,
\Gamma(\Lambda_b \to \Sigma_c\pi) = \Gamma(\Lambda_b \to \Sigma_c^*\rho) \,/\,
\Gamma(\Lambda_b \to \Sigma_c\rho) = 2 + {\cal O}[\lqcd/m_Q,\, \alpha_s(m_Q)]$.
We also discussed properties of pentaquarks with a $\bar b$ or $\bar c$,
including a possible discovery channel if they decay weakly.

\acknowledgments

We thank Shin-Shan Yu for asking questions that raised our interest in some of
these topics, and Marjorie Shapiro and Dan Pirjol for helpful discussions.
This work was supported in part by the National Science Foundation under Grant
No.~PHY-0244599 (A.K.L.); by the Director, Office of Science, Office of High
Energy and Nuclear Physics, Division of High Energy Physics, of the U.S.\
Department of Energy under Contract DE-AC03-76SF00098 (Z.L.); by the Department
of Energy under cooperative research agreement DF-FC02-94ER40818 (I.W.S); and
by the Department of Energy under Grant No.~DE-FG03-92-ER40701 (M.B.W). 
Z.L.~and I.W.S.\ were also supported by DOE Outstanding Junior Investigator
awards.

\end{document}